\documentclass[showpacs,preprintnumbers,amsmath,amssymb]{revtex4}
\usepackage{amsmath,amssymb,mathrsfs,graphics,epsfig,subfigure}
\usepackage{color}
\usepackage{float}
\usepackage{multirow}
\usepackage{booktabs}

\usepackage{enumitem}

\begin{document}
	\newcommand {\nn} {\nonumber}
	\renewcommand{\baselinestretch}{1.3}

	\title{Imprints of black hole charge on the precessing jet nozzle of M87*}
	
	%\date{\today}
	\author{Xiang-Cheng Meng,  Chao-Hui Wang, Shao-Wen Wei \footnote{Corresponding author. E-mail: weishw@lzu.edu.cn}}

	\affiliation{$^{1}$Lanzhou Center for Theoretical Physics, Key Laboratory of Theoretical Physics of Gansu Province, and Key Laboratory of Quantum Theory and Applications of MoE, Lanzhou University, Lanzhou, Gansu 730000, People's Republic of China,\\
 $^{2}$Institute of Theoretical Physics, Research Center of Gravitation, and School of Physical Science and Technology, Lanzhou University, Lanzhou 730000, People's Republic of China}

\begin{abstract}
The observed jet precession period of approximately 11 years for M87* strongly suggests the presence of a supermassive rotating black hole with a tilted accretion disk at the center of the galaxy. By modeling the motion of the tilted accretion disk particle with the spherical orbits around a Kerr-Newman black hole, we study the effect of charge on the observation of the precession period, thereby exploring the potential of this strong-gravity observation in constraining multiple black hole parameters. Firstly, we study the spherical orbits around a Kerr-Newman black hole and find that their precession periods increase with the charge. Secondly, we utilize the observed M87* jet precession period to constrain the relationship between the spin, charge, and warp radius, specifically detailing the correlations between each pair of these three quantities. Moreover, to further refine constraints on the charge, we explore the negative correlation between the maximum warp radius and charge. A significant result shows that the gap between the maximum warp radii of the prograde and retrograde orbits decrease with the black hole charge. If the warp radius is provided by other observations, different constraints on the charge can be derived for the prograde and retrograde cases. These results suggest that in the era of multi-messenger astronomy, such strong-gravity observation of precessing jet nozzle presents a promising avenue for constraining black hole parameters.
\end{abstract}


	
\maketitle
	
\section{Introduction}
\label{secIntroduction}
	
	As a prediction of general relativity (GR), the existence of black holes has been supported by the observation of LIGO \cite{LIGO}. It is widely believed that there exists supermassive black hole in the center of each galaxy. Excitingly, the Event Horizon Telescope (EHT) collaboration released the first-ever image of the black hole at the center of the M87* \cite{Akiyama1}, strongly suggesting the presence of a supermassive black hole at the center of M87 and providing a pathway for strong field tests of gravity. Subsequently, the EHT collaboration published the image of SgrA* at the center of our Milky Way \cite{Akiyama2}, further advancing the progression of the observation and theory. However, due to limitations in image resolution, there remains significant room for further studies.
	
	One of the most striking features of M87* is the bright jet of energy and matter emanating from its core. Previous studies of the inner region of M87* indicated that the jet near the black hole exhibits a large opening angle \cite{WJunor,KHada,RCWalker,RLu}. Recently, Cui et al. \cite{Cui} reported an analysis of 22 years of radio observations, showing that the jet's position angle varies periodically. They hypothesized that this is due to a misaligned accretion disk around a rotating black hole, leading to Lense-Thirring precession. Through their analysis, they derived a half-opening angle of the precession cone of $1.25^\circ \pm 0.18^\circ$ and a corresponding period of $11.24 \pm 0.47$ years, with a precession angular velocity of $0.56 \pm 0.02$ radians per year. This observation strongly indicates that the central black hole in M87* has a tilted accretion disk deviating from the equatorial plane. Subsequently, they examined the imprints of M87's jet precession on the black hole-accretion disk system, including the disk's size and the jet's non-collinear structure \cite{Cui2}.
	
	For a tilted accretion disk, the disk plane varies with radius. Within the innermost stable orbit (ISCO), matter particles will rapidly fall into the black hole, and thus it can generally regarded that the accretion disk starts at this orbit radius and then extends towards the distance. The inner disk typically undergoes Bardeen-Petterson alignment \cite{BardeenPetterson}, which means it usually aligns with the equatorial plane. The outer edge of the inner disk is defined by a characteristic radius known as the warp radius. Beyond the warp radius, the tilt angle of the disk relative to the equatorial plane gradually increases. The tilted accretion disk is found in a wide variety of systems, e.g., protostars, X-ray binaries, and active galactic nuclei (AGN) \cite{Papaloizou,Herrnstein,Begelman,Wijers,Chiang,Martin,Lodato,Casassus}. Modeling such complex accretion disks is difficult if one considers its internal dynamical mechanisms. However, if the characteristics of the tilted disk are grasped, one can simulate the motion of the accretion disk with simple and manageable model. For example, the quasi-periodic oscillations observed in certain astrophysical black hole systems can be explored by studying the precession of spherical orbits \cite{Zahrani}. Another interesting application was first proposed in Ref. \cite{Wei} that the observed jet precession period can be used to constrain the black hole parameters. In the study, the warp radius and black hole spin parameter are constrained based on the following three assumptions. First, the motion of the disk particles at each radial distance can be accurately described by spherical orbit with a constant radius, deviating from the equatorial plane \cite{Wilkins,Goldstein,Dymnikova,Shakura,ETeo,PRana,Kopek}. Second, the jet is assumed to originate near the warp radius and be oriented perpendicular to the accretion disk. Finally, the precession axis is considered as the axis of the black hole spin.
	
	The black holes of general relativity can be completely specified by only three parameters: their mass $M$, spin angular momentum $J$, and the electric charge $Q$ according to the ``no-hair theorem" \cite{Israel1,Israel2,Carter2,Hawking,Gravitation,Robinson}. The effects caused by the charge are minimal, making it difficult to constrain the charge through observations in weak gravitational fields \cite{Sereno,Ebina}. The EHT collaboration, using observations of black hole shadows in strong gravitational fields, constrained multiple parameters, including the charge, but only ruled out certain regions corresponding to specific physical charges \cite{Kocherlakota}. The precession of M87*'s jet presents another strong gravitational observation following black hole shadow, and we expect to use this observation to constrain multiple parameters. Although very recent study have explored a rotating black hole immersed in a Melvin magnetic field \cite{CChen}, a simpler and equally meaningful case is the Kerr-Newman black hole. One might argue that the charge of a charged black hole would quickly neutralize in the surrounding plasma, but here we do not consider the neutralization process or the electromagnetic interaction with the astrophysical environment. This assumption is consistent with that made in Refs. \cite{Kocherlakota,Tsukamoto}. Our goal is to explore the potential of using black hole jet precession to constrain multiple parameters, with a particular focus on constraining parameters other than the spin, specifically the charge $Q$ in the Kerr-Newman scenario. Besides, some accretion scenario also involve the study of charged rotating black holes \cite{Wilson,Damour,Ruffini}. Thus, here we consider the charged rotating black holes described by the Kerr-Newman solution at the center of M87* \cite{Newman}. Building on the assumptions proposed in Ref. \cite{Wei}, we model the tilted accretion disk using spherical orbits around a Kerr-Newman black hole and constrain the black hole parameters through the observed jet precession period.
	
	First, we calculate the energy and angular momentum of spherical orbits around a Kerr-Newman black hole, as well as the radii of the innermost stable spherical orbit (ISSO) and the last spherical orbit (LSO), and focus on the impact of the charge on these quantities. Next, we numerically solve the equations of motion in the $\theta$ and $\phi$ directions to obtain the precession angular velocity of the spherical orbits and investigate its dependence on the black hole parameters. Finally, based on these calculations, we derive the precession period and use the spherical orbits of the Kerr-Newman black hole to model the accretion disk of M87*, constraining the relationships between the spin, charge, and warp radius with the observed jet precession period. We also established a relationship between the maximum warp radius and the charge.
	
	Our paper is organized as follows. In Sec. \ref{secSO}, we carry out a detailed study of spherical orbits around a Kerr-Newman black hole, including its special subclasses ISSO and LSO. In Sec. \ref{secPOSO}, we further analyze the precession of spherical orbits. Then we provide some constraints on the black hole parameters using the observed jet precession period in Sec. \ref{secConstrain}. Finally, we discuss and conclude our results in Sec. \ref{secConclutions}. Here we adopt the metric convention $(-,+,+,+)$ and use geometrical units with $G=c=1$ in addition to recovering dimensionality in Sec. \ref{secConstrain}.

	\section{spherical orbits}
	\label{secSO}
	In this section, we study the properties of the spherical orbits for test particles around a Kerr-Newman black hole, including angular momentum, energy, and stability of the orbits. In addition, the ISSOs and LSOs are also examined. Our main focus is on the effect of a black hole's charge $Q$ and orbital tilt angle $\zeta$ on spherical orbits in the case of small and large black hole spin.
	
	We start with a brief review of the motion of test particles in the Kerr-Newman spacetime. In the Boyer-Linquist coordinates, the Kerr-Newman black hole reads
	\begin{eqnarray}
		ds^{2}=-\frac{\Delta}{\rho^{2}}\bigg(dt-a\sin^{2}\theta d\phi\bigg)^{2}
		+\frac{\rho^{2}}{\Delta}dr^{2}+\rho^{2}d\theta^{2}\nonumber\\
		+\frac{\sin^{2}\theta}{\rho^{2}}\bigg(adt-(r^{2}+a^{2}) d\phi\bigg)^{2},
	\end{eqnarray}
	where
	\begin{gather*}
		\rho^{2}=r^{2}+a^{2}\cos^{2}\theta,\\
		\Delta=r^{2}-2Mr+a^{2}+Q^{2}.
	\end{gather*}
	Here $a,Q,M$ represent the black hole's spin, charge, and mass. By solving $\Delta=0$, we easily obtain the radii of the black hole horizons
	\begin{eqnarray}\label{eqBHexist}
		r_{\pm}=M\pm\sqrt{M^2-a^2-Q^2}.
	\end{eqnarray}
	When the black hole exists, $M^2-a^2-Q^2\geq 0$ must be satisfied, otherwise, naked singularities are presented.
	
	For a given geometry, the geodesics of test particles are governed by the Hamilton Jacobi equation,
	\begin{equation}\label{eqHJeq}
		\frac{\partial S}{\partial\tau}=-\frac{1}{2}g^{\mu\nu}\frac{\partial S}{\partial x^{\mu}}\frac{\partial S}{\partial x^{\nu}},
	\end{equation}
	where $\tau$ is an affine parameter along the geodesics and $S$ is the Jacobi action. When $S$ is separable, it can be written as
	\begin{equation}\label{eqJacobiAction}
		S=\frac{1}{2}\delta\,\tau-E\,t+L\,\phi+S_{r}(r)+S_{\theta}(\theta),
	\end{equation}
	where the energy $E$ and the angular momentum $L$ per unit mass of the test particle are constants of motion associating with the Killing fields $\partial_t$ and $\partial_\phi$, respectively. We mainly study the motion of massive test particles by setting $\delta=1$. Substituting it into Eq. \eqref{eqHJeq}, the four equations of motion in the directions $\{t, r, \theta, \phi\}$ can be obtained
	\begin{eqnarray}
		 \label{eqt}\rho^{2}\frac{dt}{d\tau}&=&a\left(L-aE\sin^{2}\theta\right)+\frac{\left(r^{2}+a^{2}\right)\left(\left(r^{2}+a^{2}\right)E-aL\right) }{\Delta},\\
		\label{eqphi}\rho^{2}\frac{d\phi}{d\tau}&=&\left(L\csc^{2}\theta-aE \right)+\frac{a\left(\left(r^{2}+a^{2}\right)E-aL\right) }{\Delta} ,\\
		\label{eqr}\rho^{2}\frac{dr}{d\tau}&=&\pm\sqrt{\mathcal{R}(r)},\\
		\label{eqtheta}\rho^{2}\frac{d\theta}{d\tau}&=&\pm\sqrt{\Theta(\theta)},
	\end{eqnarray}
	with
	\begin{eqnarray*}
		\mathcal{R}(r)&=&\left(\left( r^{2}+a^{2}\right) E-aL\right)^{2}-\Delta(r^{2}+\mathcal{K}),\\
		\Theta(\theta)&=&-\left(aE\sin\theta-L\csc\theta\right)^{2}-a^2\cos^{2}\theta+\mathcal{K}.
	\end{eqnarray*}
	Here $\mathcal{K}$ is separation constant and called Carter constant corresponding to the Killing-Yano tensor \cite{Carter}.
	
	In this paper, we focus on spherical orbits as a special class of bound geodesics in Kerr-Newman spacetime with radial coordinate $r=const$. According to Eq. \eqref{eqtheta}, it is apparent that the $\theta$-motion of particles exhibits symmetry about $\theta=\frac{\pi}{2}$. Therefore, for an off-equatorial orbit, the $\theta$-motion will oscillate about the equatorial plane confined within the range $\left(\frac{\pi}{2}-\zeta,\ \frac{\pi}{2}+\zeta \right)$, where  $\zeta\in \left(0,\frac{\pi}{2}\right)$ is the tilt angle to the equatorial plane. Because of the particle turning back at two points $\theta=\frac{\pi}{2}\pm\zeta$, we have $\frac{d\theta}{d\tau}=0$, which gives
	\begin{equation}\label{eqcaterpar}
		\mathcal{K}=a^{2}\sin^{2}\zeta+\left(aE\cos\zeta-\frac{L}{\cos\zeta}\right)^{2}.
	\end{equation}
	For a spherical orbit, besides the constants $E$ and $L$, if the tilt angle $\zeta$ is given, the Carter constant $\mathcal{K}$ will also be determined. We then obtain $\theta$- and $\phi$-motions by solving Eqs. \eqref{eqtheta} and \eqref{eqphi}.

	\subsection{Spherical orbits}
	
	Spherical orbits were first studied by Wilkins in the case of Kerr black hole \cite{Wilkins}. Recently, in Ref. \cite{Zahrani}, spherical orbits are studied as the most fundamental components of tilted accretion disks. Ref. \cite{Wei} further analyses the energy, angular momentum and Carter constant associated with these orbits. Subsequently, Ref. \cite{Kopek} provides a detailed analysis of the relevant quantities associated with the ISSOs. For a Kerr-Newman black hole, spherical orbits have been studied in Ref. \cite{Alam}. Although they investigate spherical orbits for different purposes, a common initial step is to determine the constants of motion associated with these orbits. To calculate the period of  precession for spherical orbits, we first analyze the effect of charge $Q$ on energy $E$ and angular momentum $L$ under different values of spin parameter $a$ and tilt angle $\zeta$. Notably, the Carter constant $\mathcal{K}$ is determined by Eq. \eqref{eqcaterpar}. Moreover, to identify the orbits where particles can move stably, we examine the stability of spherical orbits with different energy and angular momentum.
	
	For a spherical orbit, $\dot{r}=\ddot{r}=0$. This leads to $\mathcal{R}(r)$ (in Eq. \eqref{eqr}) and its first derivative $\mathcal{R}^{\prime}(r)$ vanish,
	\begin{equation}
		\mathcal{R}(r)=\mathcal{R}^\prime(r)=0,
	\end{equation}
	where the prime denotes the derivative to $r$. Solving these two equations for $E$ and $L$, we obtain
	\begin{align}
		\label{eqE}E&=E\left(r,a,Q,\zeta\right),\\
		\label{eqL}L&=L\left(r,a,Q,\zeta\right).
	\end{align}
	The explicit forms of $E\left(r,a,Q,\zeta\right)$ and $L\left(r,a,Q,\zeta\right)$ are omitted for brevity. In this paper, we define black hole spin as being a positive direction. For prograde orbits, the particle has positive angular momentum $L > 0$, while for retrograde orbits, the particle takes negative angular momentum $L < 0$.
	
	\begin{figure}[!htbp]
		\centering{
			\subfigure[]{\includegraphics[width=5.8cm]{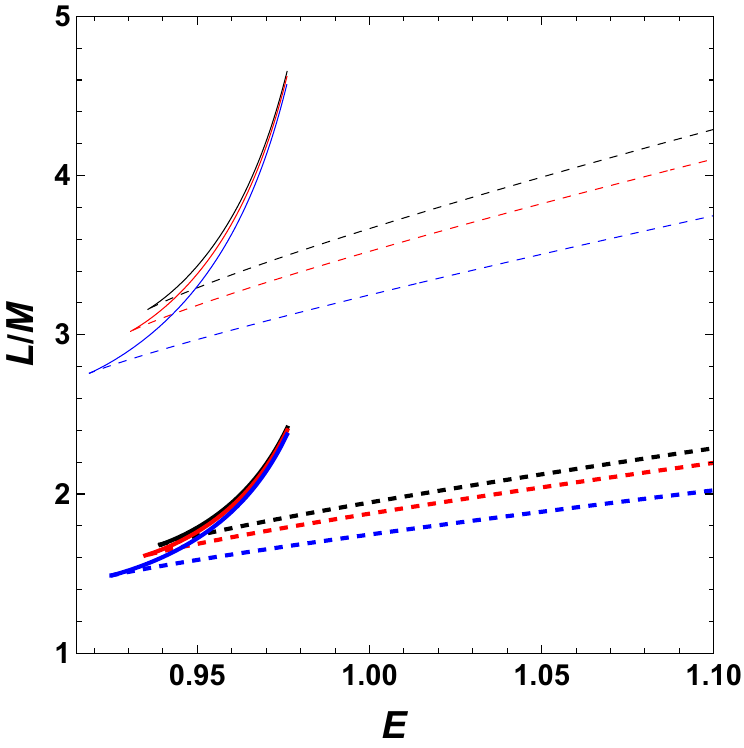}\label{figplea02}}
			\subfigure[]{\includegraphics[width=6cm]{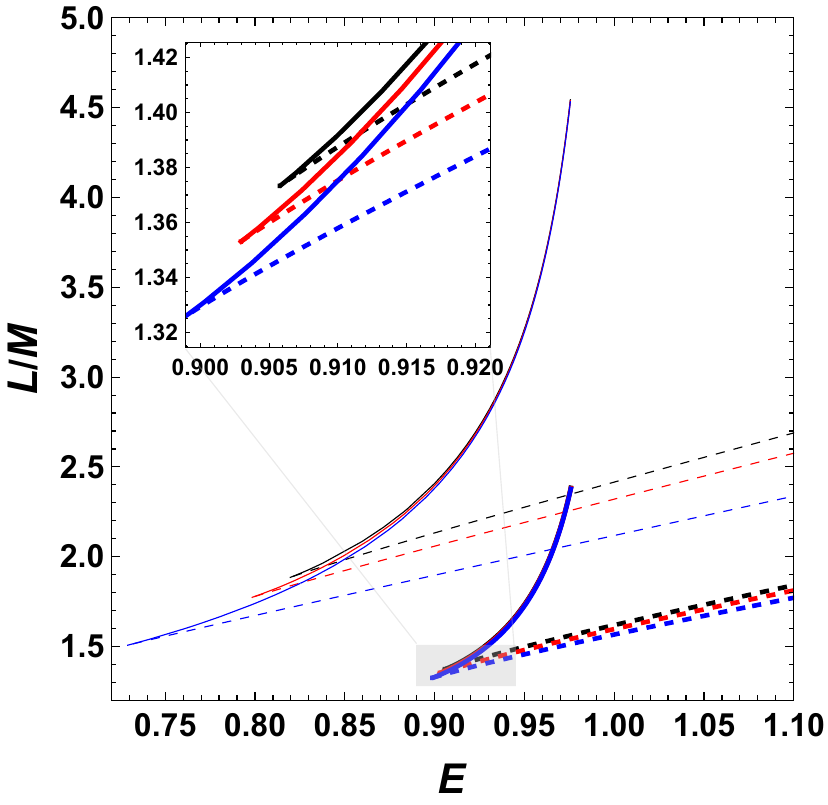}\label{figplea95}}\\
			\subfigure[]{\includegraphics[width=6cm]{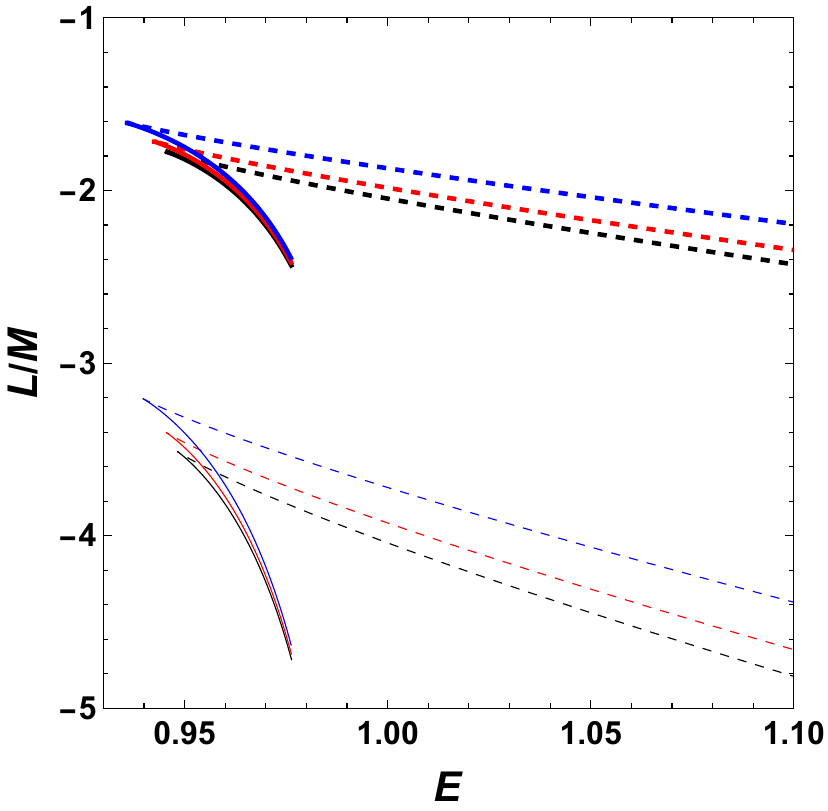}\label{fignlea02}}
			\subfigure[]{\includegraphics[width=6cm]{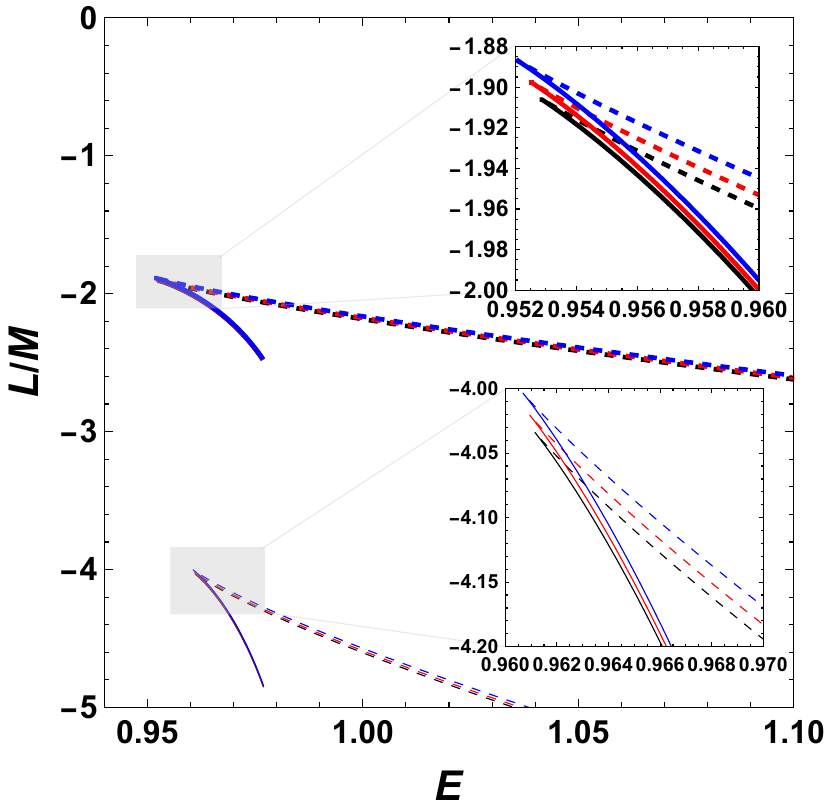}\label{fignlea95}}
			\caption{Angular momentum and energy for spherical orbits parameterized by the radius $r/M$ with $a/M=0.2, 0.95$. The upper row is for the prograde orbits and the lower row is for the retrograde orbits. The thin and thick curves correspond to $\zeta=\frac{\pi}{12}$ and $\frac{\pi}{3}$. Solid and dashed curves are for the stable and unstable spherical orbits, respectively. The intersection of the solid and dashed curves represents the ISSO. (a) $a/M=0.2$, prograde case. (b) $a/M=0.95$, prograde case. (c) $a/M=0.2$, retrograde case. (d) $a/M=0.95$, retrograde case. For the left column, the black, red and blue curves correspond to $Q/M=0, 0.5$ and $0.8$. For the right column, the black, red and blue curves correspond to $Q/M=0, 0.2$ and $0.3$. }\label{figle}
		}
	\end{figure}

	In Fig. \ref{figle}, we illustrate the relationship between the energy and angular momentum for a black hole in both high- and low-spin cases, with the radius $r$ as a parameter. Firstly, for prograde orbits, energy and angular momentum are positively correlated, while for retrograde orbits, the relationship is reversed. Secondly, in each plot, the intersection of the dashed and solid curves represents the ISSOs. Moving along the dashed curves from the ISSOs, the orbital radius gradually decreases, indicating that these orbits are unstable, as marked by the dashed curves. The absolute value of angular momentum and energy increase until they reach the LSOs, where both approach infinity. It is worth mentioning that for null geodesics ($\delta=0$ in Eq. \eqref{eqJacobiAction}), the edge of the black hole shadow corresponds to unstable spherical photon orbits \cite{Kerrshadow}, which are located closer to the event horizon. Studying these orbits can enhance our understanding of the effects of strong gravity. Conversely, moving along the solid curves from the ISSOs results in a gradual increase in orbital radius, with these orbits being stable, as marked by the solid curves. Similarly, the absolute value of angular momentum and energy also increase, approaching 1 at very large radii. For better visualization, the maximum value of $r$ is set to $20M$. We observe that the dependence of the energy and angular momentum on $r$ is similar for both high- and low-spin cases. Additionally, by varying the charge $Q$ and tilt angle $\zeta$, we find that the energy and angular momentum are more sensitive to the changes of the tilt angle. The influence of tilt angle on energy and angular momentum for stable spherical orbits is consistent with Ref. \cite{Wei}. In the high-spin case, the effect of the charge $Q$ on the energy and angular momentum is smaller due to a narrower range of $Q$ compared to the low-spin case. Notably, for stable spherical orbits at large radii, the energy and angular momentum corresponding to different values of $Q$ tend to converge. Consequently, we find that as $Q$ increases, the absolute values of energy and angular momentum decrease. The same trend is observed for the ISSOs.

	\subsection{Two special types of spherical orbits : ISSO and LSO}\label{secISSOlCO}
	In the previous subsection, we show that the stability of spherical orbits transitions from stable to unstable as the radius $r$ decreases, with the ISSOs serving as the boundary between the two types. We also introduced the LSOs, which lie within the ISSOs, where both energy and angular momentum are divergent. In this subsection, we focus on the dependence of the ISSO's and LSO's radius on the charge $Q$ and the tilt angle $\zeta$, in both high- and low-spin cases.
	
	First, we calculate the radius of the ISSOs. The stability of the spherical orbit depends on the value of $\mathcal{R}^{\prime\prime}(r)$. Specifically, a negative value of $\mathcal{R}^{\prime\prime}(r)$ corresponds to a stable spherical orbit, while a positive value corresponds to an unstable spherical orbit. For the ISSO, $\mathcal{R}^{\prime\prime}(r)$ vanishes,
	\begin{equation}
		\mathcal{R}^{\prime\prime}(r)=0.
	\end{equation}
	Substituting Eqs. \eqref{eqE} and \eqref{eqL} into it, we can solve the radius $r_{ISSO}$ .
	
	\begin{figure}[!htbp]
		\centering{
			\subfigure[]{\includegraphics[width=5cm]{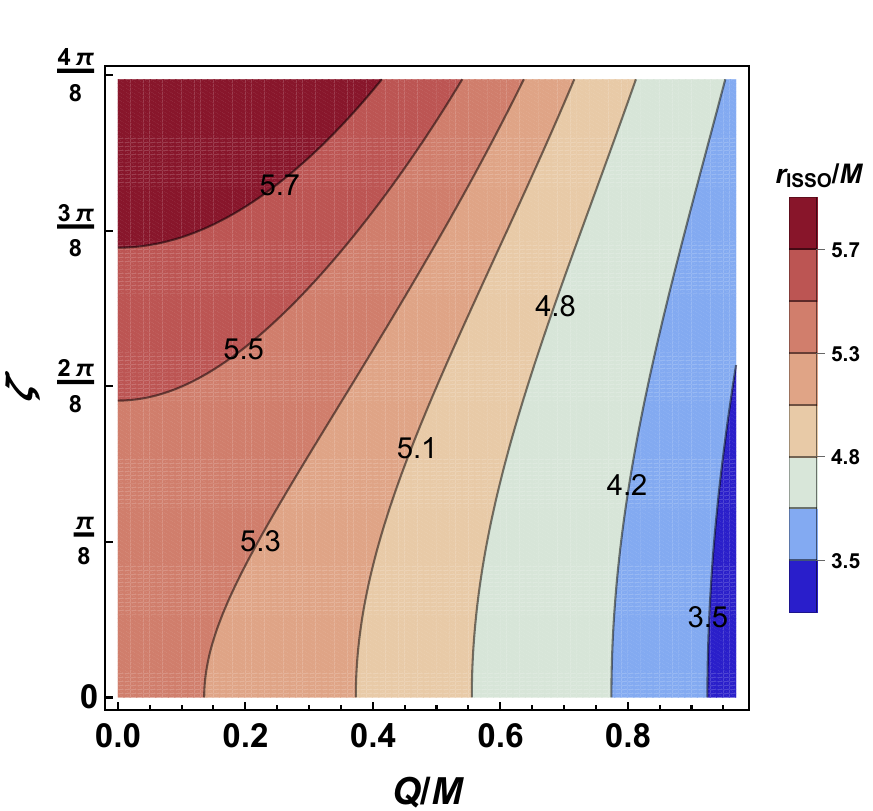}\label{figprISSOa02}}
			\subfigure[]{\includegraphics[width=5cm]{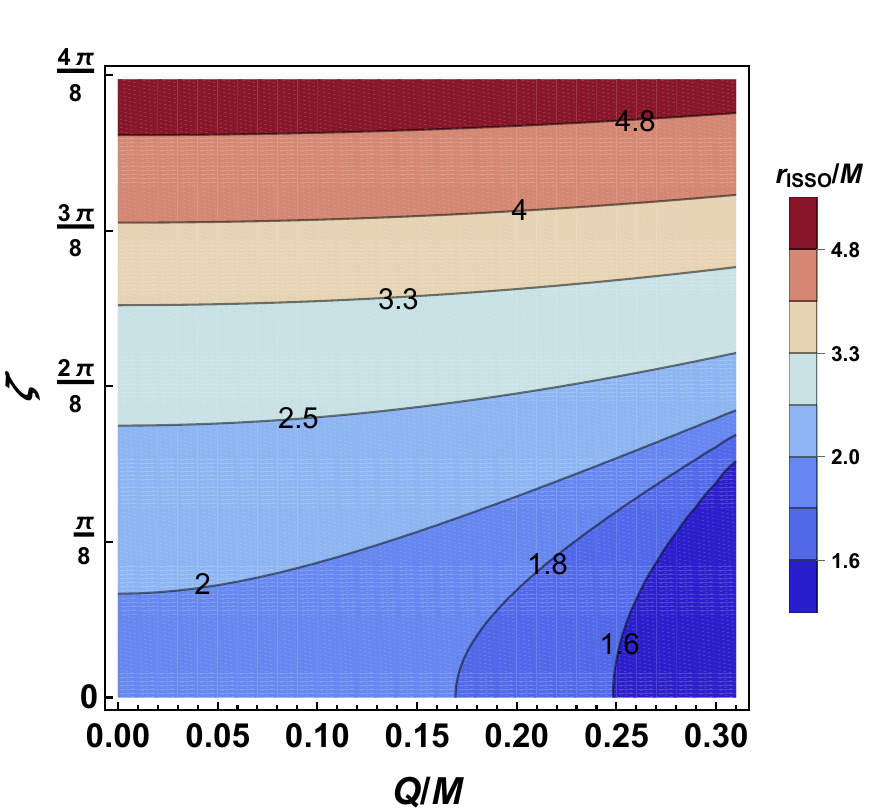}\label{figprISSOa95}}\\
			\subfigure[]{\includegraphics[width=5cm]{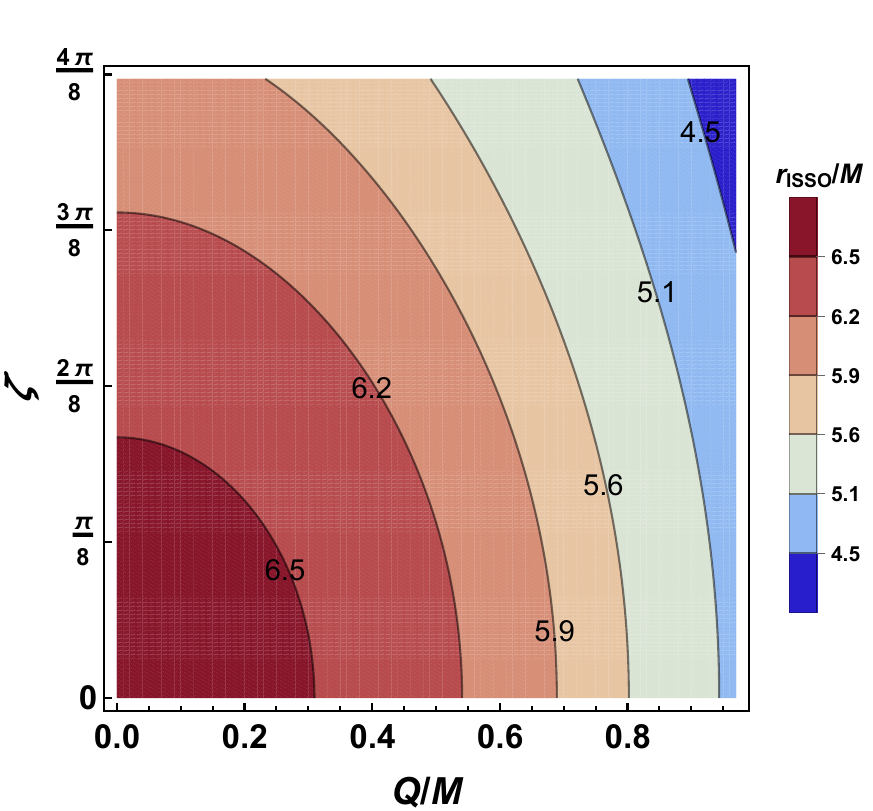}\label{fignrISSOa02}}
			\subfigure[]{\includegraphics[width=5cm]{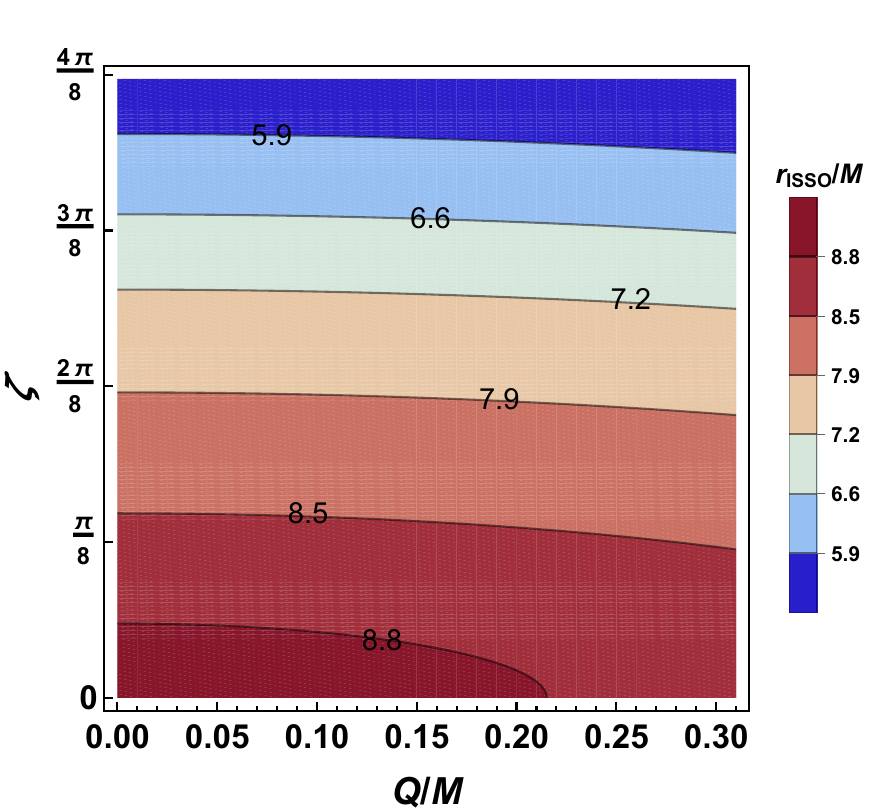}\label{fignrISSOa95}}
			\caption{Radius of the ISSO $r_{\text{ISSO}}$ as a function of the charge $Q$ and the angle $\zeta$ in the cases of low spin $a/M=0.2$ and high spin $a/M=0.95$. The upper row is for the prograde orbits and the lower row is for the retrograde orbits. (a) $a/M=0.2$, prograde case. (b) $a/M=0.95$, prograde case. (c) $a/M=0.2$, retrograde case. (d) $a/M=0.95$, retrograde case.}\label{figISSO}
		}
	\end{figure}
	Fig. \ref{figISSO} presents the variation of the radius $r_{\text{ISSO}}$ of ISSO with respect to the charge $Q$ and the tilt angle $\zeta$. We observe that $r_{\text{ISSO}}$ decreases with the charge $Q$. As $\zeta$ increases, $r_{\text{ISSO}}$ increases for prograde orbits and decreases for retrograde orbits, which is consistent with the result of Fig. 4 (a) of Ref. \cite{Wei}. Furthermore, we find that the dependence of $r_{\text{ISSO}}$ on $Q$ is more obvious in the low spin case than in the high spin case. It is worth noting that for prograde orbits, $r_{\text{ISSO}}$ is less than $6M$ in most cases, except for scenarios with low spin, low charge, and large tilt angle. In contrast, for retrograde orbits, $r_{\text{ISSO}}$ exceeds $6M$ under conditions of low charge and small tilt angle, or in the high spin case.
	
	\begin{figure}[!htbp]
		\centering{
			\subfigure[]{\includegraphics[width=5cm]{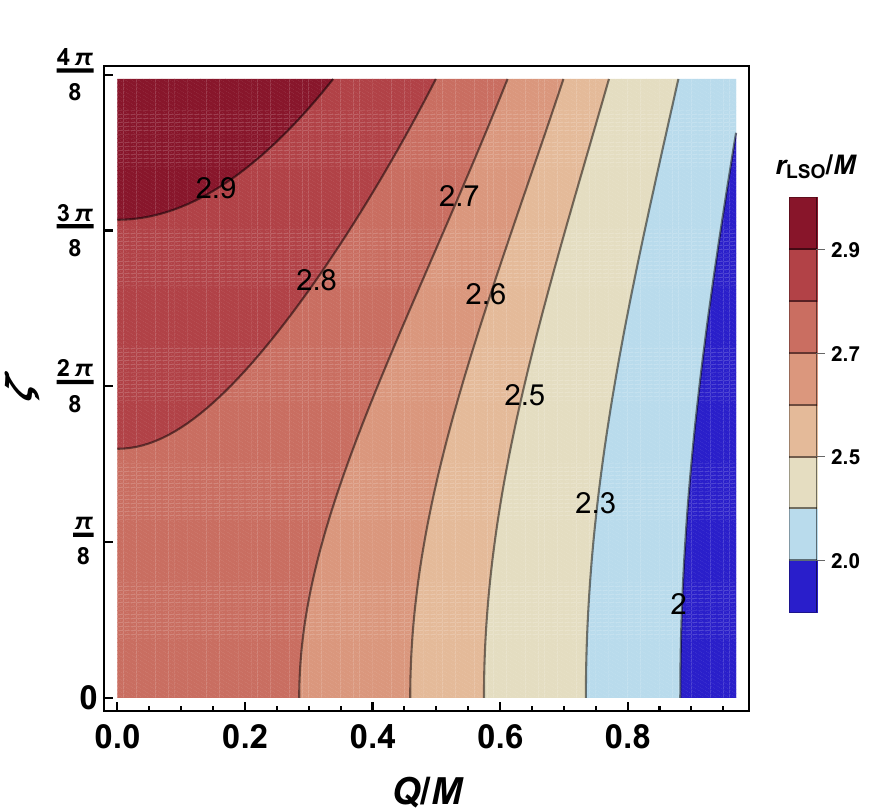}\label{figprLSOa02}}
			\subfigure[]{\includegraphics[width=5cm]{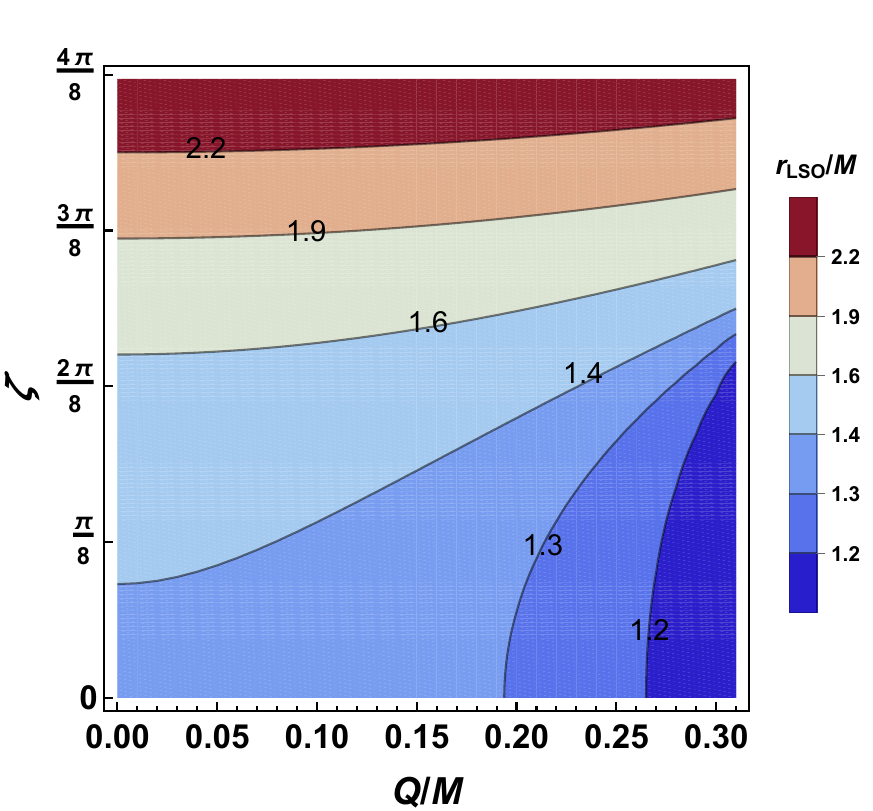}\label{figprLSOa95}}\\
			\subfigure[]{\includegraphics[width=5cm]{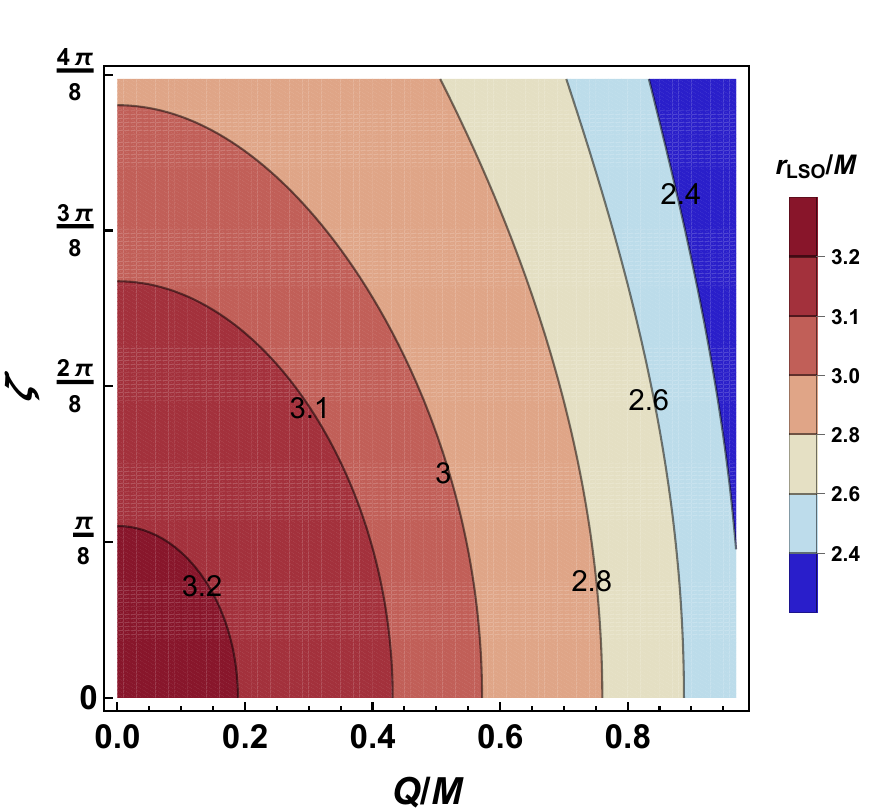}\label{fignrLSOa02}}
			\subfigure[]{\includegraphics[width=5cm]{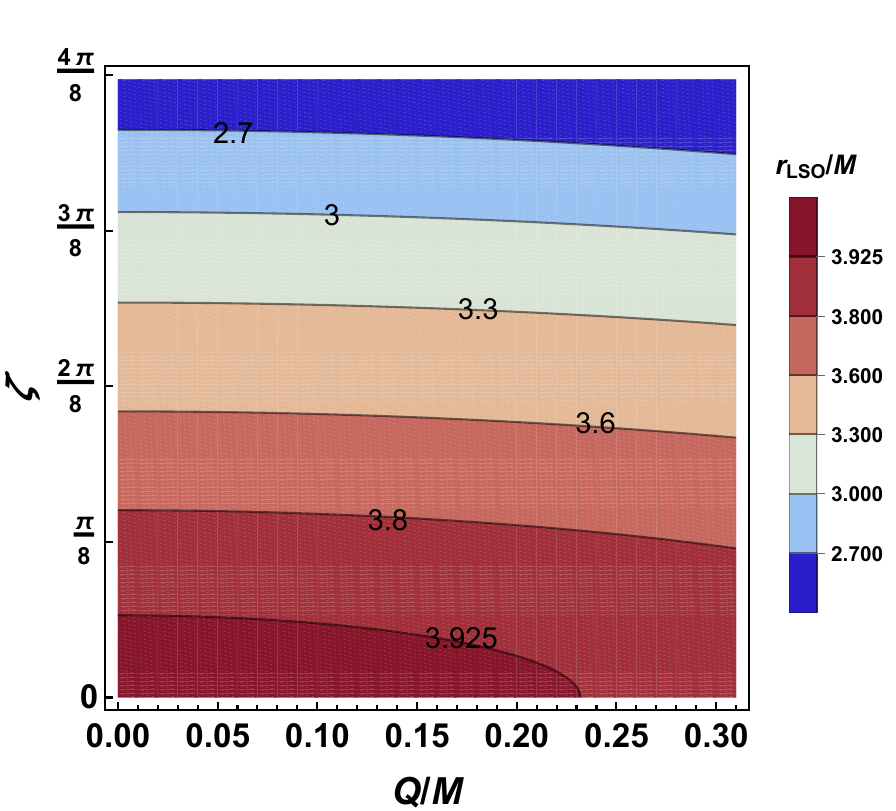}\label{fignrLSOa95}}
			\caption{Radius $r_{\text{LSO}}$ of the LSO as a function of the charge $Q$ and the angle $\zeta$ in the cases of low spin $a/M=0.2$ and high spin $a/M=0.95$. The upper row is for the prograde orbits and the lower row is for the retrograde orbits. (a) $a/M=0.2$, prograde case. (b) $a/M=0.95$, prograde case. (c) $a/M=0.2$, retrograde case. (d) $a/M=0.95$, retrograde case.}\label{figLSO}
		}
	\end{figure}
	
	For the LSO, the energy and angular momentum are divergent, which implies that its radius $r_{\text{LSO}}$ can be obtained by setting $E^{-1} = 0$. In Fig. \ref{figLSO}, we present the radius of the LSO as a function of $Q$ and $\zeta$ with $a/M=0.2$ and $0.95$. Our results indicate that the radius of the LSOs exhibits a similar dependence on both $Q$ and $\zeta$ as that observed for ISSOs. The result of Ref. \cite{Zahrani} reveals that the dependence of $r_{\text{ISSO}}$ and $r_{\text{LSO}}$ on the spin parameter $a$ are also similar. Therefore, we speculate that the influence of the black hole parameters on $r_{\text{ISSO}}$ and $r_{\text{LSO}}$ is the same. In addition, it is observed that the LSOs are in close to ISSOs when $Q$ is at its maximum, which results in a narrow interval of unstable spherical orbits between them. This phenomenon can be attributed to the strong gravity present in the near event horizon region when the black hole parameters reach their extremal values.
	
	\section{Precession of spherical orbits}\label{secPOSO}
	
	In the static spherically symmetric spacetime, if the particles are not constrained to the equatorial plane, the spherical orbits of the particles must be a tilted ring. Considering the symmetry, the spherical orbits can be cast in the equatorial plane by reselecting the coordinate axis. Differentially, in the stationary axisymmetric spacetime, the periods of the $\theta$ and $\phi$ directions of the spherical orbits of particles are different due to the dragging effect of black holes in the $\phi$ direction, resulting in Lense-Thirring precession. In this section, we focus on the precession of spherical orbits around a Kerr-Newman spacetime.
	
	Here, we consider the precession of spherical orbits in the vicinity of the black hole as seen by a distant observer, which differs from the local observer given in Ref. \cite{Zahrani}. In astrophysics, observers and astrophysical events are generally at significant distances from each other. Thus, it is necessary to parameterise the motion of $\theta$ and $\phi$ using the coordinate time $t$. From Eqs. \eqref{eqt}, \eqref{eqtheta}, and \eqref{eqphi}, we obtain
	\begin{eqnarray}
		 \label{eqdthetadt}\frac{d\theta}{dt}&=&\frac{\pm\sqrt{\Theta(\theta)}}{a\left(L-aE\sin^{2}\theta\right)+\frac{\left(r^{2}+a^{2}\right)\left(\left(r^{2}+a^{2}\right)E-aL\right) }{\Delta}},\\
		\label{eqdphidt}\frac{d\phi}{dt}&=&\frac{\left(L\csc^{2}\theta-aE \right)+\frac{a\left(\left(r^{2}+a^{2}\right)E-aL\right) }{\Delta}}{a\left(L-aE\sin^{2}\theta\right)+\frac{\left(r^{2}+a^{2}\right)\left(\left(r^{2}+a^{2}\right)E-aL\right) }{\Delta}},
	\end{eqnarray}
	where $E$ and $L$ are solved in Eqs. \eqref{eqE} and \eqref{eqL}.
	
	By numerically integrating these two equations, we can obtain the motion of $\theta$ and $\phi$ directions with initial conditions $\theta (0) = \frac{\pi}{2}$ and $\phi (0) = 0$. From the evolution curves of $\theta$ and $\phi$ with respect to the coordinate time $t$, we can extract the precession angular velocity of the motion in the $\phi$ direction relative to that in the $\theta$ direction. For specific calculation details, we refer to Ref. \cite{Wei}.
	
	Now, let us calculate the precession angular velocity $\omega_{t}$ of the spherical orbits. Integrating Eq. \eqref{eqdthetadt}, we obtain the period of $\theta$ motion $T_{\theta}$. From the $\phi$ motion, we can obtain the change $\Delta \phi$ within time $T_{\theta}$. Thus, the precession angular velocity $\omega_{t}$ is given by
	\begin{eqnarray}
		\label{eqomega}\omega_{t}=\frac{\Delta\phi-2\pi}{T_{\theta}}.
	\end{eqnarray}
	
	\begin{figure}[!htbp]
		\centering{
			\subfigure[]{\includegraphics[width=5cm]{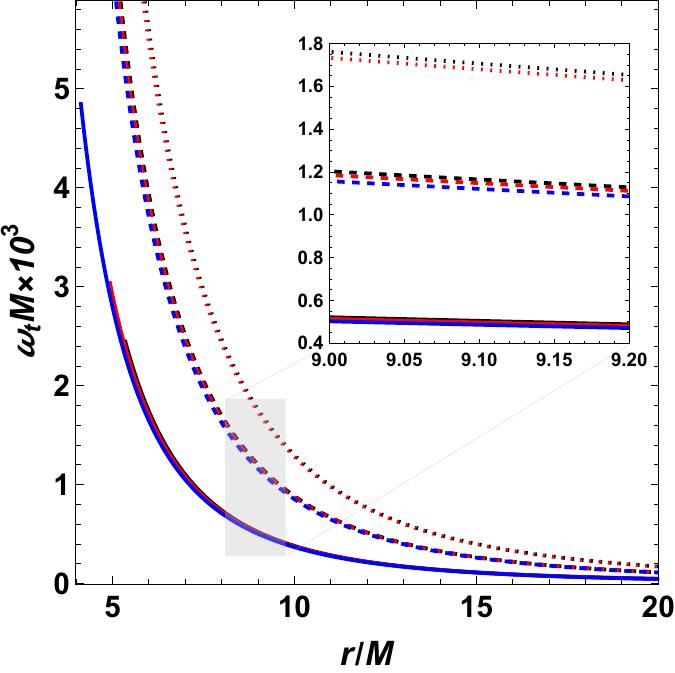}\label{figpomegatpi12}}
			\subfigure[]{\includegraphics[width=5cm]{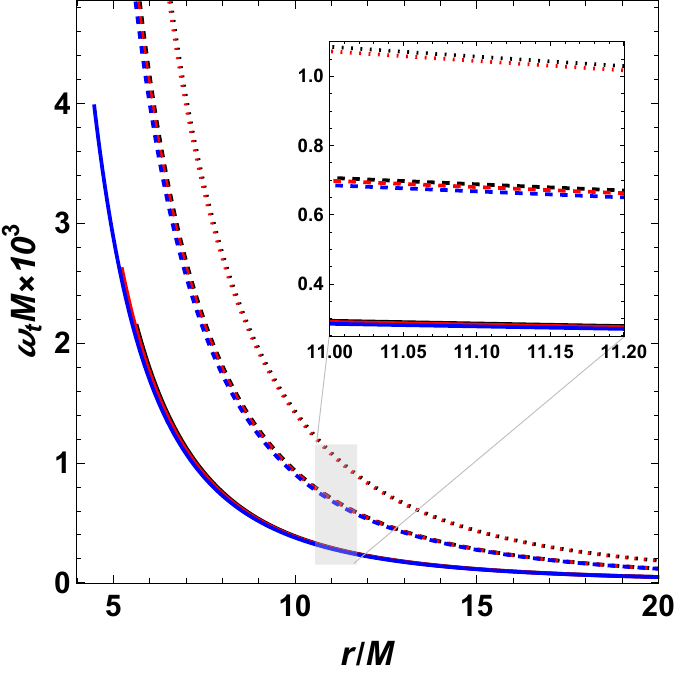}\label{figpomegatpi3}}\\
			\subfigure[]{\includegraphics[width=5cm]{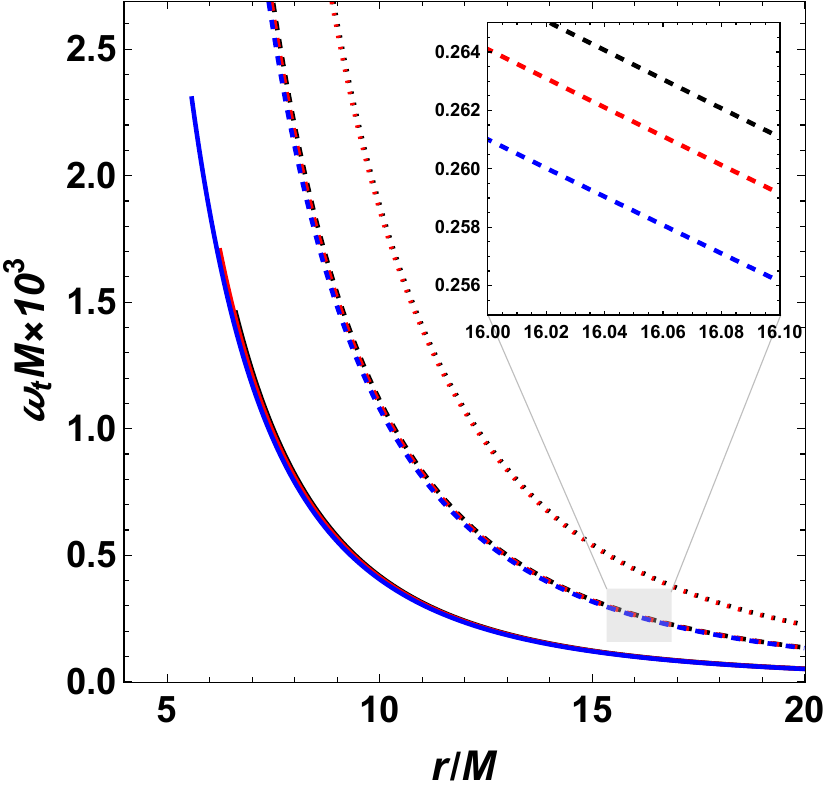}\label{fignomegatpi12}}
			\subfigure[]{\includegraphics[width=5cm]{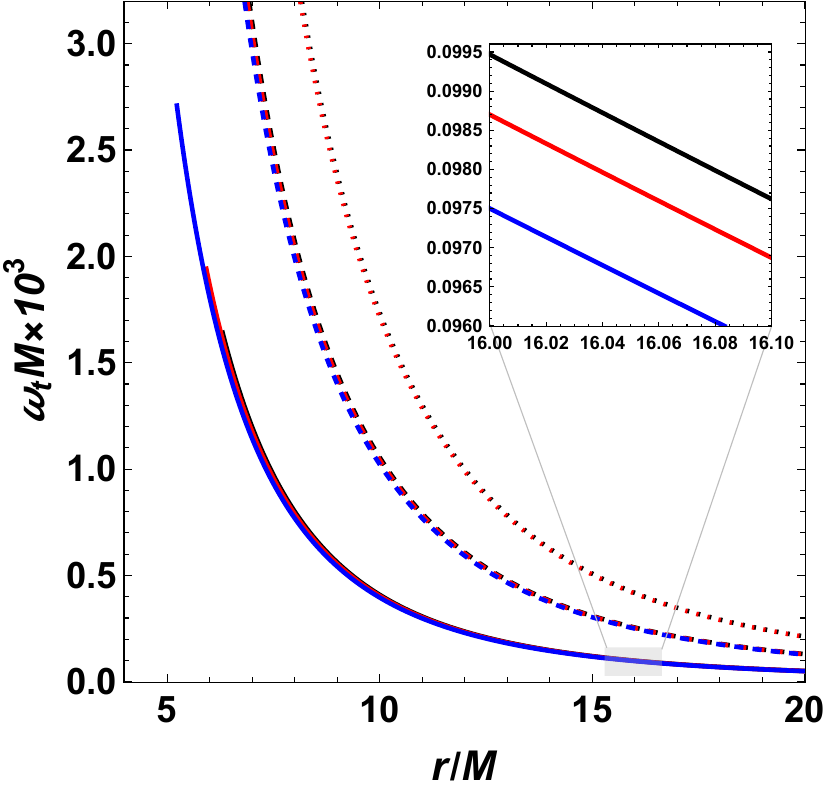}\label{fignomegatpi3}}
			\caption{The precession angular velocity $\omega_{t}$ as a function of radius $r$, ranging from $r_{ISSO}$ to $20M$, with $\zeta = \pi/12$ and $\pi/3$. The left column is for the prograde orbit and the right column is for the retrograde orbit. The upper row is for the prograde orbits and the lower row is for the retrograde orbits. The solid, dashed and dotted curves correspond to $a/M=0, 0.5$, and $0.8$, respectively. The black, red and blue correspond to $Q/M=0, 0.5$, and $0.8$, respectively. (a) $\zeta=\pi/12$, prograde case. (b) $\zeta=\pi/3$, prograde case. (c) $\zeta=\pi/12$, retrograde case. (d) $\zeta=\pi/3$, retrograde case.}\label{figomegat}
		}
	\end{figure}
	
	The numerical results in Ref. \cite{Zahrani} indicate that variations in the tilt angle $\zeta$ have a negligible effect on the precession angular velocity $\omega_t$, despite their significant impact on the energy and angular momentum as analyzed earlier. Fig. \ref{figomegat} also shows minimal differences between the left and right subplots. In astronomical observations, the tilt angle is usually known, and fixing it is helpful for constraining parameters using the precession period in the subsequent analysis. We can see that the effect of changing $Q$ on $\omega_{t}$ is small compared to changing $a$. This is because the necessary condition for precession mainly comes from the spin $a$. If the angular velocity $\omega_{t}$ is expanded as a series in terms of $a$ and $Q$, the leading order contains only $a$, and $Q$ disappears (corresponding to the Kerr case). Furthermore, the metric is a function of $Q^{2}$. Therefore, the influence of $Q$ is significantly smaller than that of $a$. We can infer that for a compact object, even if its charge changes, the precession angular velocity of the surrounding spherical orbits will not experience significant changes. Interestingly, the dependence of the angular velocity $\omega_t$ on $a$ and $Q$ is different: $\omega_t$ increases with $a$, while decreases with $Q$. We also note that regardless of whether the particle is in a prograde or retrograde orbit, the precession angular velocity always aligns with the direction of the black hole's spin, reflecting the dragging effect of the rotating black hole. Finally, it is clear that as the radius $r$ increases, $\omega_t$ consistently decreases. This suggests that the greater the distance from the black hole, the weaker the dragging effect becomes.
	
	\section{Precession period and Constrains of M87*}\label{secConstrain}
	
	Recently, Cui et al. \cite{Cui} reported a period of approximately $11$ years for the variation in the position angle of the jet based on the analysis of radio observation of galaxy M87 over $22$ years. They infered that they are seeing a spinning black hole which occurs in the Lense-Thirring precession of a misaligned accretion disk. Subsequently, this observational fact was first used to constrain the spin of the black hole by using the precession of spherical orbits \cite{Wei}. In previous work \cite{Petterson,Ostriker,Fragile,LodatoPrice}, a clear picture has been developed for a tilted accretion disks. As the radial distance decreases, the tilt angle of the disk decreases. At a characteristic radius known as the warp radius, the disk returns to the equatorial plane. When the particles are inside the ISSO, they will rapidly fall into the black hole. Thus, the warp radius can exceed the radius of the ISSO. In complex astrophysical environments, simulating an accretion disk requires taking into account many factors, such as disk viscosity, magnetic fields, external forces or torques, and so on. This task is typically handled by magnetohydrodynamics. However, we focus on the most fundamental components of the accretion disk, using spherical orbits to model the motion of particles within the disk. Additionally, combining this with the image of a tilted disk, we propose that jets originate near the warp radius, leading to a small misalignment between the jet axis and the black hole's spin axis. General relativistic magnetohydrodynamic (GRMHD) simulations have demonstrated that a significant portion of the accretion disk in these misaligned systems undergoes Lense-Thirring precession \cite{PCFragileOBlaes,Liska,White,Chatterjee,Ressler}, and that the jet precesses in sync with the disk \cite{Liska,JCMcKinney}. Under our assumption, the precession period of the jet is naturally consistent with the precession period of the spherical orbits near the warp radius. The purpose of these assumptions is to extract as much information as possible about the black hole parameters, using the observed precession of the jet of M87* black hole. Although these assumptions are overly simplified, they provide a fast and effective way to establish relationships between black hole parameters, serving as a preliminary step for more precise simulations.
	
	To use the precession period to constrain black hole parameters, we first summarize how to calculate the precession period for spherical orbits. The precession period after unit restoration, is given
	\begin{eqnarray}
		\label{eqperiod}T=\frac{2\pi}{\omega_{t}}\frac{GM_{\odot}}{c^{3}}\left(\frac{M}{M_{\odot}}\right)\approx 9.80244\times10^{-13}\times\frac{1}{\omega_{t}}\left(\frac{M}{M_{\odot}}\right)\left(year\right),
	\end{eqnarray}
	where $M_{\odot}$ is the mass of sun. From the observation, the mass of M87$^{\star}$ black hole is $M=6.5\times 10^{9}M_{\odot}$ \cite{Akiyama1}. From Eqs. \eqref{eqdthetadt} and \eqref{eqdphidt}, we know that the functional form of $\omega_{t}$ is given by
	\begin{equation}
		\omega_{t}=f(r, a, Q, E(r, a, Q, \zeta), L(r, a, Q, \zeta))
	\end{equation}
	According to Ref. \cite{Cui}, the half-opening angle of the precession cone is $1.25^\circ \pm 0.18^\circ$. However, Ref. \cite{Zahrani} points out that the effect of changing $\zeta$ on $\omega_t$ is negligible, so we fix the tilt angle at $\zeta_{p}=1.25^\circ$. It is worth noting that current observations do not clearly determine the angular momentum direction of the accretion disk, making it essential to distinguish between prograde and retrograde orbits in the calculations. Given $a$, $Q$, and $r$, we first obtain the energy $E(r, a, Q, \zeta_{p})$ and angular momentum $L(r, a, Q, \zeta_{p})$ following the calculations in Sec. \ref{secSO}. Then, we compute the angular velocity $\omega_{t}$ by following Sec. \ref{secPOSO}. Finally, by substituting it into Eq. \eqref{eqperiod}, we obtain the precession period $T$. The ranges of $a$ and $Q$ are determined by the existence conditions of the black hole, as described in Eq. \eqref{eqBHexist}. For $r$, we are particularly interested in the radius associated with the origin of the jets, namely the warp radius, since its corresponding precession is linked to the precession of the jets. In Ref. \cite{Zahrani}, the range for the warp radius is set between (6$M$, 20$M$). From Fig. \ref{figISSO}, we observe that in many cases, $r_{ISSO}$ is less than 6$M$. To comprehensively consider all spherical orbits, we set the radius range to $(r_{ISSO}, 20M)$. In the subsequent analysis, we will find that the maximum constrained value of the warp radius is always less than 20$M$.
	
	In summary, the precession period $T$ of the jets relative to the black hole spin axis is determined by the black hole spin $a$, charge $Q$, and the warp radius $r$. In Ref. \cite{Cui}, it is reported the precession period of the M87* jet as $11.24 \pm 0.47$ years. This constrains a family of surfaces in the $\{a, Q, r\}$ parameter space, where $T$ is constant on each surface. Our task is to establish the relationship between these three parameters. The relation between $a$ and $r$ is shown in Fig. 7 of Ref. \cite{Wei}, where $Q=0$ (Kerr black hole). If $Q$ takes other values, a similar relationship holds.
	
	\subsection{Relationship between $a$, $Q$ and $r$}
	
\begin{figure}[!htbp]
		\centering{
			\subfigure[]{\includegraphics[width=4.9cm]{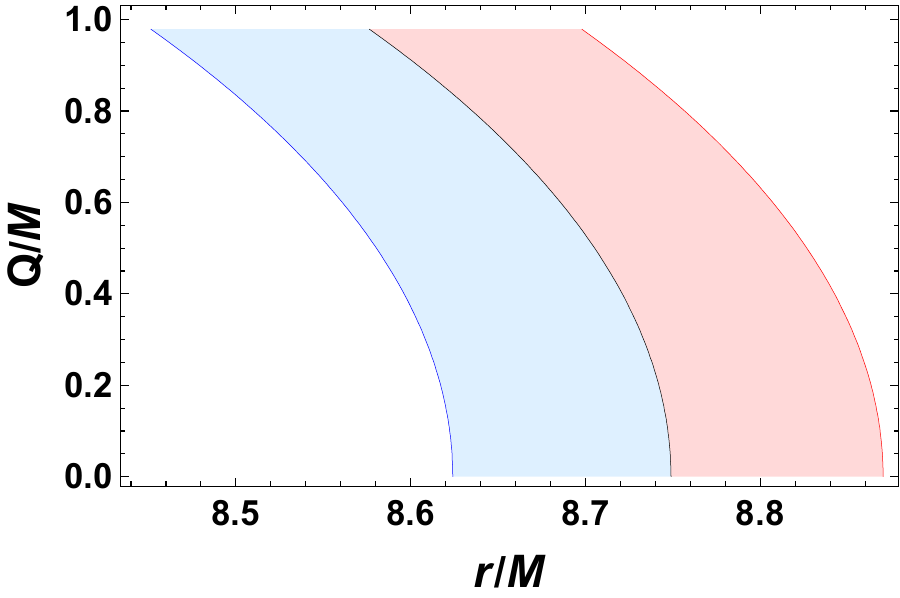}\label{figpQra02}}
			\subfigure[]{\includegraphics[width=5cm]{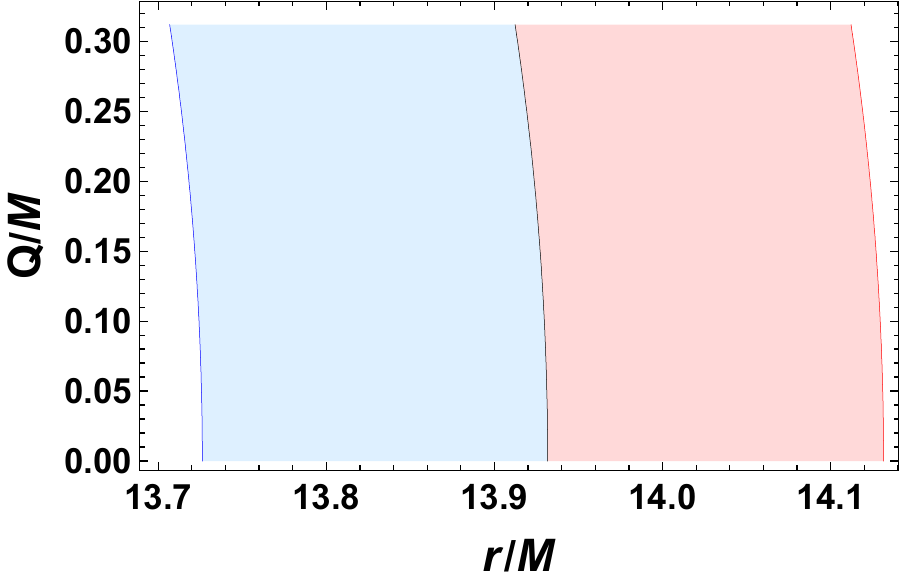}\label{figpQra95}}\\
			\subfigure[]{\includegraphics[width=4.9cm]{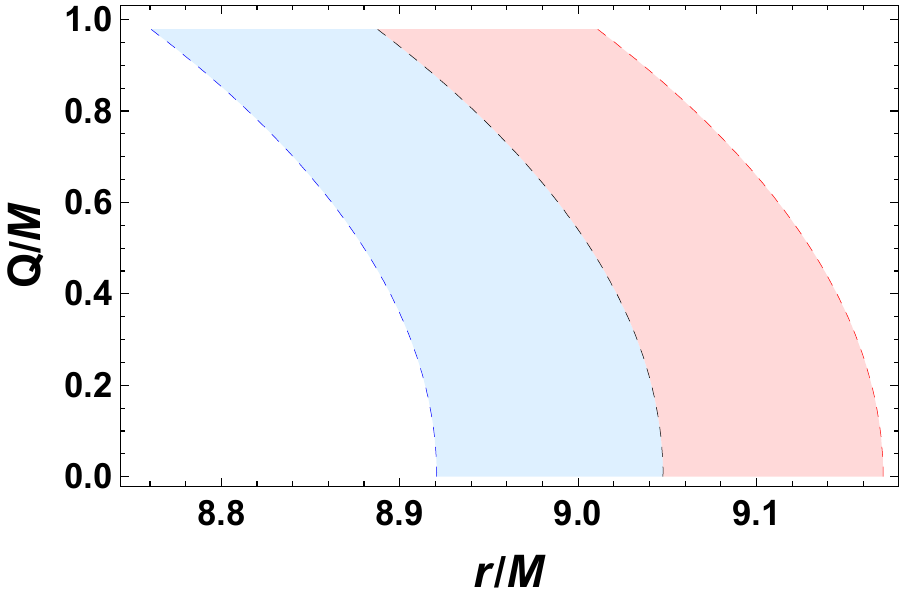}\label{fignQra02}}
			\subfigure[]{\includegraphics[width=5cm]{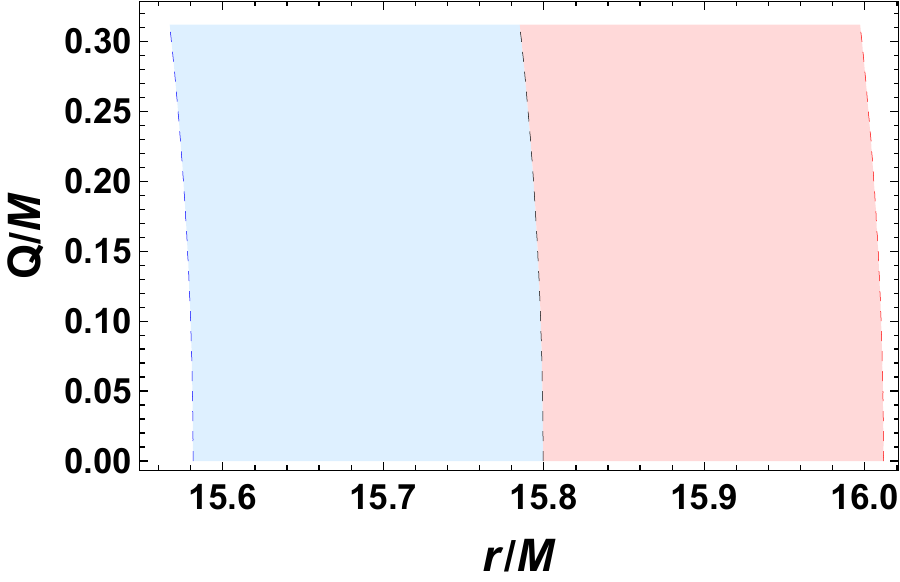}\label{fignQra95}}
			\caption{Constrain the charge $Q$ and warp radius $r$ of the accretion disk by the observed precession period of the jet nozzle of M87$^{*}$. The prograde orbits are represented by solid thin curves above, while the retrograde orbits are depicted with dashed thin curves below. The black, red and blue curves correspond to $T=11.24, 11.71$ and $10.77$ years, respectively. (a) $a/M=0.2$, prograde case. (b) $a/M=0.95$, prograde case. (c) $a/M=0.2$, retrograde case. (d) $a/M=0.95$, retrograde case.}\label{figQr}
		}
	\end{figure}

	Fig. \ref{figQr} shows the constraints on the charge $Q$ and warp radius $r$ of the accretion disk with $a/M=0.2$ and $0.95$ by the observed precession period of the jet nozzle of	M87$^{*}$. We find that the charge decreases with the warp radius for fixed $a$. The warp radius is constrained to a narrow range, although our calculations range from $r_{ISSO}$ to $20M$ for the warp radius. This is because the dependence of the angular velocity on $a$ is much more sensitive compared to $Q$, as shown in Fig. \ref{figomegat}. The same applies to the precession period from Eq. \eqref{eqperiod}. Here we fix $a$, which results in a narrow range for the warp radius. We also observe subtle differences in the relationship between $Q$ and $r$ for high-spin and low-spin cases, with the curve being more tortuous in the low spin. Additionally, we observe that retrograde orbits have a larger warp radius compared to the prograde ones. For better constraint, the observed precession period must be measured more precisely, so that the shaded region in the figure could become narrower.

\begin{figure}[!htbp]
		\centering{
			\subfigure[]{\includegraphics[width=5cm]{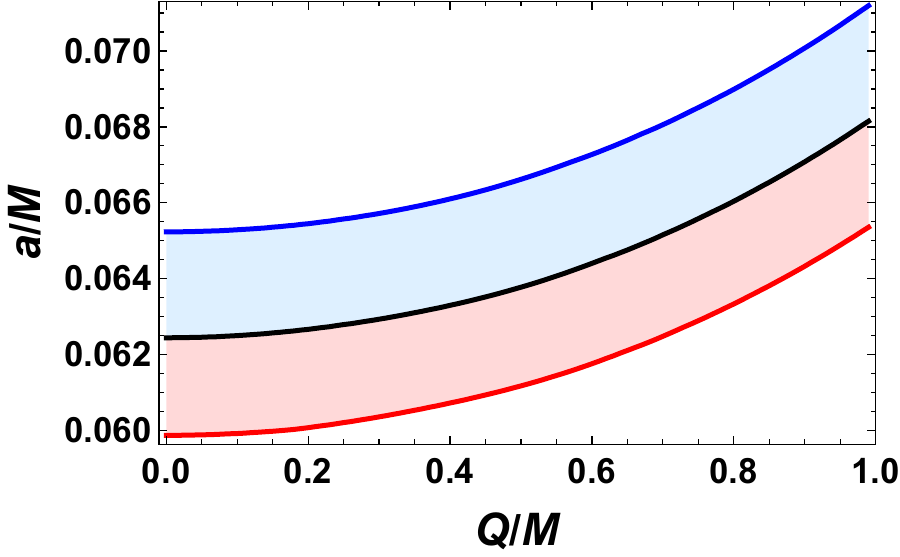}\label{figpaQr6}}
			\subfigure[]{\includegraphics[width=4.8cm]{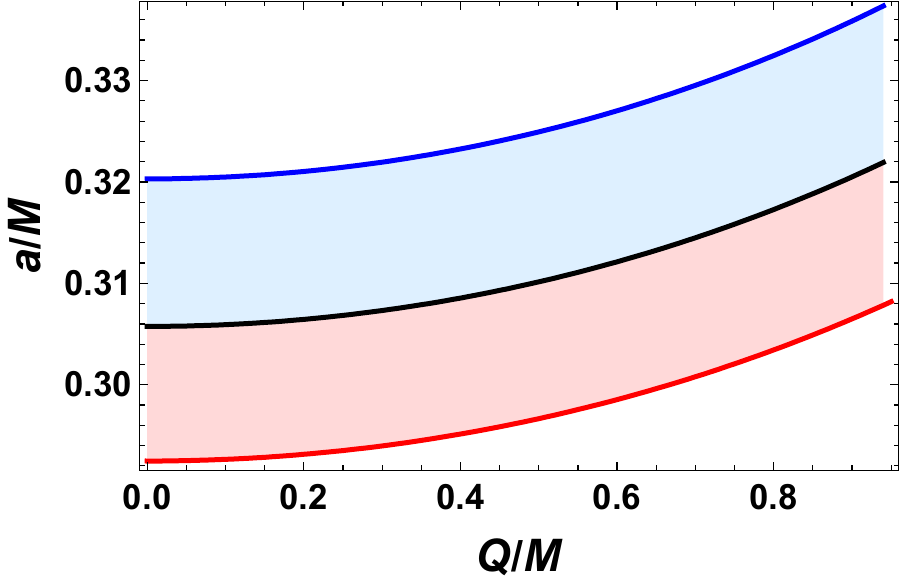}\label{figpaQr10}}\\
			\subfigure[]{\includegraphics[width=5cm]{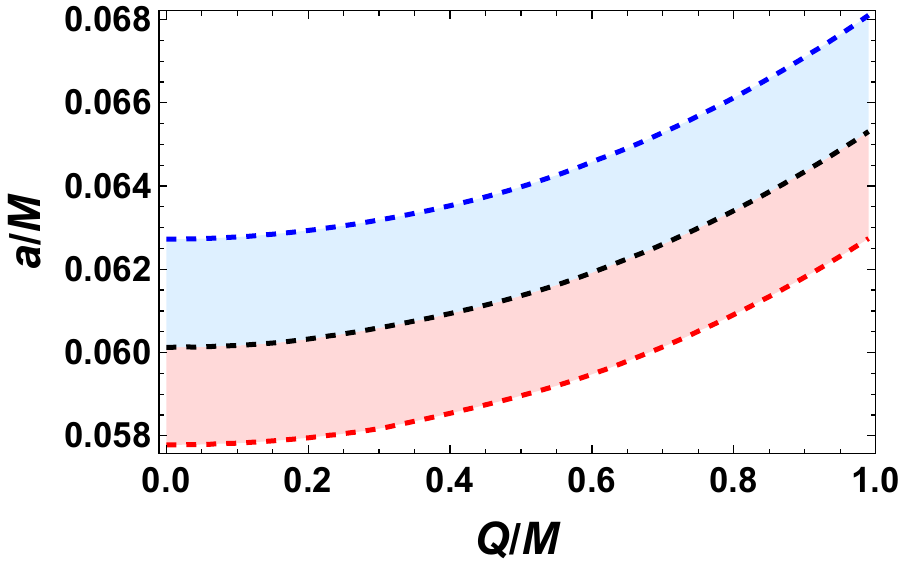}\label{fignaQr6}}
			\subfigure[]{\includegraphics[width=4.9cm]{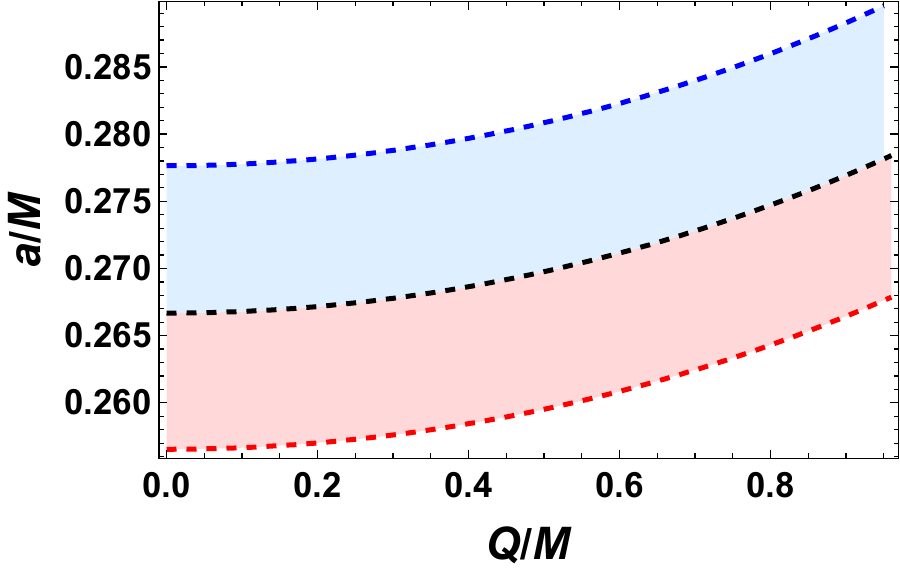}\label{fignaQr10}}
			\caption{Constrain the spin parameter $a$ and the charge $Q$ by the observed precession period of the jet nozzle of M87$^{*}$. The prograde orbits are represented by solid thick curves above, while the retrograde orbits are depicted with dashed thick curves below. The black, red and blue curves correspond to $T=11.24, 11.71$ and $10.77$ years, respectively. (a) $r/M=6$, prograde case. (b) $r/M=10$, prograde case. (c) $r/M=6$, retrograde case. (d) $r/M=10$, retrograde case.}\label{figaQ}
		}
	\end{figure}

	In Fig. \ref{figaQ}, we present the relationship between $a$ and $Q$ with fixed $r$ from the observation. Clearly, the spin parameter increases slowly with the charge for fixed $r$. The effect of varying the warp radius on the spin parameter $a$ is more pronounced. For smaller warp radius, spin $a$ for prograde and retrograde orbits is very close, but as $r$ increases, the difference becomes more noticeable, with the spin being larger in the prograde case.
	
	\subsection{Maximum values of the warp radius and charge}
	
	Due to the multiple parameters and observational precision issues, the results above seem rough, only providing constraints between the parameters. To further tighten these constraints, we calculated the relationship between the maximum warp radius and the charge. Ref. \cite{Wei} concluded that for a fixed precession period, the warp radius and spin of a Kerr black hole ($Q = 0$) are positively correlated. For $Q \neq 0$, this conclusion can be drawn by comparing the values of $a$ and $r$ on both sides of Fig. \ref{figaQ}. Thus, for each fixed charge $Q$, when the black hole spin reaches its maximum, the warp radius also reaches its maximum value.
	
\begin{figure}[!htbp]
		\centering{
			\includegraphics[width=10cm]{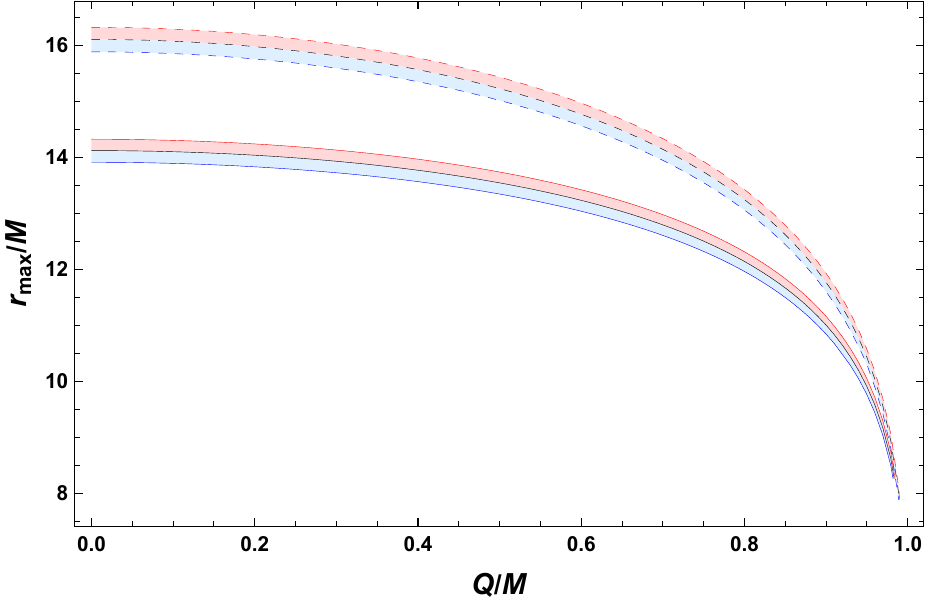}
			\caption{The relationship between the maximum value of warp radius of the accretion disk $r_{max}$ and charge $Q$ is derived from the observed precession period of the jet nozzle of M87$^{*}$. The prograde orbits are represented by solid thin curves, while the retrograde orbits are depicted with dashed thin curves. The black, red and blue curves correspond to $T=11.24, 11.71$ and $10.77$ years, respectively.}\label{figrmax}
		}
	\end{figure}

	In Fig. \ref{figrmax}, we show that the maximum warp radius $r_{max}$ is negatively correlated with the charge $Q$. Additionally, the maximum warp radius for retrograde orbits is generally larger than the one for prograde orbits. As $Q$ increases, the constrained region becomes narrower, and as it approaches $M$, the difference between them becomes very small. We provide serval data in Table \ref{t1}. Here, we find that the maximum warp radius for a Kerr black hole is $14.12^{+0.20}_{-0.21}M$ for prograde orbits and $16.11^{+0.22}_{-0.22}M$ for retrograde orbits, where the uncertainty range (represented by the "+" and "-") corresponds to the observational constraints on the precession period, with an upper bound of 11.71 years and a lower bound of 10.77 years. This constraint is more precise than that of Ref. \cite{Wei}.
	
	Importantly, we observe from the figure that when $r_{\text{max}}$ lies between $14.12M$ and $16.11M$, only retrograde orbits exist. This is a significant result, as it could provide critical insights into the black hole of M87*. For instance, if future observations measure a warp radius of $15M$, we can infer that the accretion disk is rotating retrograde relative to the black hole spin axis and constrain the black hole's charge to $Q \lesssim 0.55M$. However, if the observed warp radius is in the range $(8M,14.12M)$, we would not be able to distinguish between prograde and retrograde orbits. Naturally, each case imposes different constraints on the charge, with prograde orbits offering stronger constraints. For example, if the observed warp radius is $13M$, the charge is limited to $Q \lesssim 0.66M$ for prograde orbits and $Q \lesssim 0.82M$ for retrograde orbits. While these results are based on our simplified toy model, more accurate values would require detailed numerical simulations that take into account the complexities of astrophysical environments. Nonetheless, our model outlines general trends, offering guidance for precise numerical simulations and contributing insights to theoretical studies of black hole accretion disks and jet models.

	\begin{table}[h]
		\centering
		
		\addtolength\tabcolsep{6pt}
		{\color{black}
			\begin{tabular}{c|cc}
				\hline    \hline
				\multirow{2}*{$Q/M$} &	\multicolumn{2}{c}{$r_{\text{max}}/M$} \\  \cline{2-3}
				&	prograde &	retrograde\\
				\hline \hline
				0	&	$14.12^{+0.20}_{-0.21}$	&	$16.11^{+0.22}_{-0.22}$	\\
				0.1	&	$14.10^{+0.20}_{-0.21}$	&	$16.07^{+0.22}_{-0.22}$	\\
				0.2	&	$14.04^{+0.20}_{-0.21}$	&	$15.98^{+0.21}_{-0.22}$	\\
				0.3	&	$13.93^{+0.20}_{-0.21}$	&	$15.81^{+0.21}_{-0.22}$	\\
				0.4	&	$13.77^{+0.20}_{-0.20}$	&	$15.57^{+0.21}_{-0.22}$	\\
				0.5	&	$13.54^{+0.19}_{-0.20}$	&	$15.23^{+0.20}_{-0.21}$	\\
				0.6	&	$13.23^{+0.19}_{-0.19}$	&	$14.77^{+0.20}_{-0.21}$	\\
				0.7	&	$12.79^{+0.18}_{-0.19}$	&	$14.14^{+0.19}_{-0.20}$	\\
				0.8	&	$12.14^{+0.17}_{-0.18}$	&	$13.24^{+0.18}_{-0.19}$	\\
				0.9	&	$11.00^{+0.16}_{-0.16}$	&	$11.76^{+0.16}_{-0.17}$	\\
				\hline
		\end{tabular}}
		\caption{For a given charge $Q$, the maximum warp radius $r_{\text{max}}$ is provided for both prograde and retrograde cases. The upper and lower bounds of each $r_{\text{max}}$ correspond to the constraints from the upper and lower limits of the precession period $T$, respectively.
		}
		\label{t1}
	\end{table}
	
	\section{Discussions and conclusions}\label{secConclutions}
	
	In this paper, we focused on the spherical orbits of test particles around a Kerr-Newman black hole and constrained the black hole parameter by using the recent observation of the precessing jet nozzle of M87*. Following the observation of the black hole's shadow, this observation has emerged as another observational tool in strong gravitational fields, providing a promising method to constrain the properties of supermassive black holes and test gravitational theories. The periodic precession of the jet from the supermassive black hole at the center of M87*, relative to the black hole's spin axis, suggests that the jet originates from a tilted accretion disk. The tilt of the disk induces a kinematic precession of particles within the disk relative to the spin axis. This simplified physical picture motivates us to use spherical orbits around a black hole to model the tilted accretion disk and thus determine the jet precession period closely associated with the disk.
	
	To achieve this, we first analyzed a subclass of bound orbits, specifically spherical orbits with constant radius. In addition to the energy and angular momentum, the motion constants of these orbits include the Carter constant $\mathcal{K}$, which arises from the separability of the geodesics and is determined by fixing the tilt angle relative to the equatorial plane. We found that, for various values of spin $a$ and tilt angle $\zeta$, as the charge $Q$ increases, the absolute value of angular momentum $L$ and energy $E$ decrease. The distinction between the prograde and retrograde orbits lies in the different correlation between energy and angular momentum: for prograde orbits, they are positively correlated, while for retrograde orbits, the correlation is negative. The above results apply to ISSO as well. Significantly, near the LSO, the energy and angular momentum exhibit divergent behavior. We further investigated the radial distribution of spherical orbits, and found that it is primarily determined by the ISSO and LSO, the two special spherical orbits. As the charge $Q$ increases, both $r_{\text{ISSO}}$ and $r_{\text{LSO}}$ decrease.
	
	After thoroughly studying the properties of spherical orbits, we solved for the energy and momentum of these orbits, which serves as the foundation for our further analysis. By numerically solving the geodesic equations in the $\theta$ and $\phi$ directions, it is found that the motion in the $\theta$ direction is periodic, while the motion in the $\phi$ direction undergoes precession relative to the $\theta$-motion. Then, we calculated the precession angular velocity $\omega_t$ as seen by a distant observer and found that its dependence on the charge $Q$ is much weaker compared to the spin parameter $a$. The angular velocity $\omega_t$ increases with $a$, while decreases with $Q$ and $r$. Furthermore, regardless of whether the orbit is prograde or retrograde, the direction of the angular velocity is always aligned with the black hole's spin, reflecting the frame-dragging effect of the rotating black hole.
	
	Finally, we obtained the precession period $T$ from Eq. \eqref{eqperiod}. For each set of values of $r$, $a$, and $Q$, we repeated the above steps to obtain the corresponding period. Observations show that the jet from the M87* black hole forms an angle of $1.25^\circ \pm 0.18^\circ$ with the spin axis, with a precession period of $11.24 \pm 0.47$ years. We assumed that the black hole's jet originates near the warp radius, which marks the boundary between the region of the disk that is off the equatorial plane and the region that remains aligned with it. To account for all possible spherical orbits, we considered the warp radius $r$ within the range $(r_{ISSO}, 20M)$ and calculate the precession period for each point in the $r$, $a$, and $Q$ parameter space. Consequently, the observed precession period constrains a family of surfaces in this parameter space, with each surface corresponding to values in the range $10.77$ to $11.71$ years. We further calculated the parameter relationships on these constrained surfaces: for a fixed spin $a$, the charge $Q$ decreases as the warp radius $r$ increases, and the warp radius for prograde orbits is smaller than that for retrograde orbits. For fixed warp radius $r$, the spin $a$ increases slowly with the charge $Q$, and at larger warp radii, the difference between prograde and retrograde orbits becomes more pronounced. Although these constraints do not provide definitive parameter values, they qualitatively limit the correlations between the parameters. To further constrain the charge $Q$, we investigated the relationship between the maximum warp radius $r_{max}$ and charge $Q$, revealing a negative correlation between $r_{max}$ and $Q$. We found that the gap between the maximum warp radius for prograde and retrograde orbits decreases as the charge increases. If future observations determine that the warp radius lies between $14.12M$ and $16.11M$, the accretion disk can be confirmed to be counter-rotating relative to the black hole's spin axis, allowing an upper limit on the charge to be set. If the warp radius falls between $8M$ and $14.12M$, it will not be possible to distinguish between prograde and retrograde orbits, but different upper limits on the charge can be obtained, with the retrograde case allowing for a larger upper limit. If the warp radius is less than $8M$, this method cannot provide strict constraints on the charge.
	
	In summary, our calculations and analysis offer a method to constrain black hole parameters, especially providing constraints on the black hole's charge in certain cases. While our assumptions are relatively simple, and the specific numerical values require more accurate simulations, our qualitative conclusions are significant and offer a reference for future precise calculations. As multi-messenger astronomy progresses, combining different observational methods and data may offer new opportunities for constraining black hole parameters using jet precession periods.

	\section*{Acknowledgements}
	This work was supported by the National Natural Science Foundation of China (Grants No. 12075103, No. 12475055, and No. 12247101).
	
	\bibliographystyle{unsrt}

\end{document}